
\input jnl
\input eqnorder
\input reforder
\def\ad{a^{\dag}}
\def\phib{\bar\phi}
\def\const{{\rm const.}}
\def\el{\ell}
\def\nt{\tilde n}
\def\oxon{All Souls College\\{\rm and}\\Department of Physics\\
Theoretical Physics\\1 Keble Road\\Oxford OX1 3NP, UK}
\def\refs#1{Refs.\thinspace#1}
\def\ffrac#1#2{\textstyle{#1\over#2}\displaystyle}
\def\preprintno#1{\rightline {\rm #1}}
\preprintno{OUTP--94--35S}
\title
Proportion of Unaffected Sites in a Reaction-Diffusion Process
\author
John Cardy
\affil\oxon
\abstract
We consider the probability $P(t)$ that a given site remains
unvisited by any of a set of random walkers in $d$ dimensions
undergoing the reaction $A+A\to0$ when they meet. We find that
asymptotically $P(t)\sim t^{-\theta}$ with a universal exponent
$\theta=\ffrac12-O(\epsilon)$ for $d=2-\epsilon$, while, for $d>2$,
$\theta$ is non-universal and depends on the reaction rate. The analysis,
which uses field-theoretic renormalisation group methods, is also applied to
the reaction $kA\to0$ with $k>2$. In this case, a stretched exponential
behaviour is found for all $d\geq1$, except in the case $k=3$, $d=1$,
where $P(t)\sim {\rm e}^{-\const (\ln t)^{3/2}}$.

\endtitlepage

In a recent letter, Derrida, Bray and Godr\`eche\refto{DBG}
have found new non-trivial
and apparently universal
exponents associated with the zero-temperature relaxational dynamics of
the one-dimensional Ising and Potts models. The occurrence of such
exponents is surprising given the trivial nature of the conventional
static and dynamic exponents in these models. However, the quantity
considered by these authors, namely the probability $P(t)$
that, starting from
a random initial configuration, a given site has not been crossed by a
domain wall, is not simply related to the usual response functions, and
might be expected to show more interesting behaviour. The fact that
a simple universal power law $P(t)\sim t^{-\theta}$ is obtained for large $t$,
however, requires some explanation.

In this letter, we provide such an explanation within the context of a
generalisation of this problem to arbitrary dimensionality $d$. Since
the motion of domain walls for $d>1$ is very difficult to treat
analytically, instead we observe that, in one dimension,
the motion and annihilation
of Ising domain walls at zero temperature is equivalent to a
reaction-diffusion process of point particles $A$ undergoing the
irreversible reaction $A+A\to0$. The study of this problem is readily
capable of generalisation to arbitrary $d$, and many of its features
have already been elucidated using a number of approaches.\refto{A} In
particular, it is found that there is an upper critical dimension
$d_c=2$ above which the mean density $n(t)$ behaves as $1/(\lambda t)$,
where $\lambda$ is the reaction rate, as
predicted by a simple rate equation neglecting correlation effects,
while for $d<2$ these effects cannot be ignored and the behaviour is
modified to $t^{-d/2}$, with an amplitude independent of $\lambda$.
Recently, a systematic field-theoretic
renormalisation group approach to this problem has been
developed,\refto{BLEE}
which not only yields the exponents but also correlation functions and
universal amplitudes within a $\epsilon$-expansion.
It is also straightforward to generalise the analysis to the reaction $kA\to0$.
In this case, above the upper critical dimension $d_c(k)=2/(k-1)$
one finds $n(t)\sim 1/(\lambda t)^{d_c(k)/2}$, while for $d<d_c$,
$n(t)\sim t^{-d/2}$ with a universal amplitude.

Within this type of reaction-diffusion problem, then, we ask the
following question: from a random initial condition (mean density
$n_0$) at time $t=0$, what is the late time dependence of the
probability $P(t)$ that a given site has never been visited by a
walker?
A simple approach to this problem is to note that $P(t)-P(t+\delta t)$
is the probability of a finding a walker at the given site (the origin, say),
in the time interval $(t,t+\delta t)$, given that the origin has never been
visited in the past. This will happen only if a particle happens to lie close
to
the origin at time $t$. Thus
$$
-P'(t)\delta t\sim D\delta t\,P(t)\,\nt(t)
$$
where $D$ is the diffusion constant, and $\nt(t)$ is the density at a site
adjacent to the origin, given that the origin is never visited, that is,
with a repulsive potential there.

For $d>2$, only a finite fraction of particles near the origin has ever
visited the
origin in the past, so that $\nt(t)\propto n(t)\sim t^{-d_c(k)/2}$.
For $k=2$, this leads to $P(t)\sim t^{-\const/\lambda}$, that is, a
power law with
a {\it non-universal} exponent, while for $k>2$ we obtain a
stretched exponential
behaviour $P(t)\sim {\rm e}^{-\const t^{(k-2)/(k-1)}}$.
For $d<2$, however, almost all particles
near the origin have visited it at some time in the past,
so that $\nt(t)\ll n(t)$.
When $2>d>d_c(k)$ (which is possible when $k>2$),
particle correlations may be neglected
and we may use the inhomogeneous rate equation
$$
\partial n/\partial t=D\nabla^2n-k\lambda n^k\quad,
\eqno(RE)
$$
selecting the required events by imposing the condition
$n(r=0,t)=0$. This problem has the
radially symmetric scaling solution $n(r,t)=t^{-d_c(k)/2}f(r/(Dt)^{1/2})$.
For $r\ll(Dt)^{1/2}$,
the nonlinear term is unimportant (corresponding to the fact
that the density is so low that
annihilation events rarely occur), and $f$ satisfies Laplace's equation,
with solution
$f\sim(r/(Dt)^{1/2})^\epsilon$, where $\epsilon=2-d$.
Thus $\nt(t)\sim n(t)t^{-\epsilon/2}$,
and the stretched exponential becomes
$P(t)\sim {\rm e}^{-\const t^{(d-d_c(k))/2}}$. For
$d<d_c(k)$, if we assume that $\nt(t)$ is suppressed
relative to the bulk density by this
same factor,
we find that $\nt(t)\sim t^{-d/2-\epsilon/2}=t^{-1}$, resulting in a power
law $P(t)\sim t^{-\theta}$, consistent with the result of
Derrida {\it et al.}.\refto{DBG}
However, to justify this argument and to demonstrate the
universality of $\theta$, it
is necessary to proceed more systematically.

We first relate $P(t)$ to an appropriate
correlation function in the field-theoretic description of the problem.
We follow the notation and formalism of \ref{BLEE}. Following
Doi\refto{DOI} and Peliti\refto{PEL}, the master equation for the
reaction-diffusion problem is encoded in a hamiltonian, or liouvillean,
which may expressed in the `second-quantised' form
$$
H=(D/b^2)\sum_{\rm n.n.}(\ad_i-\ad_j)(a_i-a_j)-\lambda
\sum_i(1-(\ad_i)^k)a_i^k
$$
where the first term is a sum over nearest neighbours, and represents
a continuous-time random walk on a lattice with spacing $b$, and the
second term represents the annihilation process $kA\to0$. The time
translation operator is ${\rm e}^{-Ht}$, which may be written as a
path-integral by dividing the interval $(0,t)$ into small slices of
duration $\Delta t$. At each slice a complete set of coherent states
$$
\int {\rm e}^{-\phi^*_t\phi_t}{\rm e}^{\phi_t\ad}|0\rangle\langle0|{\rm
e}^{\phi^*_ta}
d\phi^*_td\phi_t
$$
is inserted (lattice labels are suppressed for clarity).
The matrix elements\break
$\langle0|{\rm e}^{\phi^*_{t+\Delta t}a}
{\rm e}^{\phi_t\ad}|0\rangle={\rm e}^{\phi^*_{t+\Delta t}\phi_t}$ then give
rise, when
combined with the measure factors ${\rm e}^{-\phi^*_t\phi_t}$, to the
time-derivative piece in the action
$$
S=\int(\phi(t)^*\partial_t\phi(t)+D\nabla\phi^*\cdot\nabla\phi
+\lambda({\phi^*}^k-1)\phi^k)dt
$$
in the limit $\Delta t\to0$.

In the second quantised formalism, the probability $P(t)$ that a given site,
which may chosen to be the origin $j=0$,
is never visited is simply given by inserting the operator
$\delta_{\ad_0a_0,0}$ at each time slice. Thus, at the origin, instead
of the matrix element given above, one should compute
$$
\langle0|{\rm e}^{\phi^*_{t+\Delta t}a}\,\delta_{\ad a,0}\,{\rm
e}^{\phi_ta}|0\rangle=1
$$
which means that, to leading order in $\Delta t$, the factor from the
measure is not cancelled, corresponding to an insertion
of $\prod_t{\rm e}^{-\phi^*_{0,t}\phi_{0,t}}$ in the path integral. In the
continuum limit the lattice fields are rescaled so that
$\phi^*\phi\to b^d\phi^*\phi$, so that finally
$$
P(t)=\int{\cal D}\phi^*{\cal D}\phi\,{\rm e}^{-h\int\phi^*(0,t)\phi(0,t)dt}
\,{\rm e}^{-S[\phi^*,\phi]}
$$
where $h=b^d/\Delta t$.

As explained in \refs{3,5}, in order to compute statistical
averages with respect to the action $S$, it is necessary to evaluate
projections onto the state $\langle0|\prod_j{\rm e}^{a_j}$.
It is then convenient to make
the shift $\phi^*=1+\phib$, so that $\phib$ annihilates this state when
acting to the left. The required insertion then has two pieces
$$
{\rm e}^{-h\int_0^t\phib(0,t')\phi(0,t')dt'}\quad
{\rm e}^{-h\int_0^t\phi(0,t')dt'}
$$
In the first factor, the time integral can be taken up to infinity,
and thus the term $h\int\phib(0,t')\phi(0,t')dt'$ may be regarded as
a repulsive potential and included in the action $S$,
while the second factor is the piece whose
expectation value we now wish to evaluate with respect to this modified
action. It is convenient to rewrite this as a cumulant expansion
$$
P(t)\sim\langle {\rm e}^{-h\int_0^t\phi(0,t')dt'}\rangle
={\rm e}^{-h\int_0^t\langle\phi(0,t')\rangle dt'+
h^2\int_0^t\int_0^{t'}\langle\phi(0,t')\phi(0,t'')\rangle_cdt'dt''+\cdots}
\eqno(CUM)
$$

Dimensional analysis then dictates that $h$ has dimension
$({\rm wave\ number})^{2-d}$, so the additional interaction term
is {\it irrelevant}
for $d>2$, and may therefore be neglected in studying the late time
asymptotics. For $d>d_c(k)$, the arguments of \ref{BLEE} show that
all loop corrections to the field theory are also irrelevant. The sum of
the tree diagrams is then given by the solution to the naive rate
equation, $\langle\phi(0,t)\rangle\sim1/(\lambda t)^{d_c(k)/2}$.
(In fact the amplitude will be modified in the neighbourhood of the
origin, but not the exponent, for $d>2$.)
On substitution into the first term of the cumulant expansion, this then
leads to the same result as the earlier naive argument. The higher order
terms in the cumulant expansion all involve at least one more power of
$\lambda$, and hence their integrals are down by successive powers of
$t^{-(d-d_c(k))/2}$.
The appearance of the borderline dimensionality
$d=2$ is simply related to the recurrence property of random walks.

For $d<2$, however, the $h$ interaction is either marginal or relevant,
and it is necessary to perform a full renormalisation group analysis.
Fortunately this is fairly simple since the renormalisations of $h$
and $\lambda$ do not mix. This is because the renormalisation of $h$ may be
discussed in terms of its contribution to the propagator
$\langle\phi(x,t)\phib(x',t')\rangle$, to which $\lambda$ does not
contribute, while, since $\lambda$ is a bulk coupling, its renormalisation
cannot be affected by the localised interaction $h\phib\phi(0)$. The
bulk renormalisation of $\lambda$ is discussed in \ref{BLEE}. Here we use
the notation $\el_R$ to denote the dimensionless renormalised coupling
$\lambda_R\kappa^{-2\epsilon'/d_c(k)}$, where $\epsilon'=d_c(k)-d$.
The renormalisation of $h$ is then carried out for $\lambda=0$. Since
this is a Gaussian theory, this is simple, but non-trivial due to the
localised form of the interaction. The renormalised coupling $h_R$ may
be defined in terms of the truncated Fourier-Laplace transform of
$\langle\phi(x,t)\phib(0,t')\phi(0,t')\phib(x'',t'')\rangle$,
evaluated at the normalisation point imaginary frequency $s=\kappa^2$.
We thus find
$$
h_R=h\left(1+hI_d(\kappa)\right)^{-1}
$$
where
$$
I_d(\kappa)=\int{1\over s+p^2}{d^dp\over(2\pi)^d}\bigg|_{s=\kappa^2}
\equiv B(d){\kappa^{-\epsilon}\over\epsilon}
$$
The dimensionless coupling $g_R=h_R\kappa^{-\epsilon}$ (where now
$\epsilon=2-d$) then has the beta function
$$
\beta_g(g_R)=-\epsilon g_R+B(d)g_R^2
$$
where $B(2)=1/2\pi$. This is exact to all orders in $g_R$. In addition
to the coupling constant renormalisation, however, it is necessary to
perform a multiplicative renormalisation of $\phi(0,t)$. This may be
seen by considering the correlation function
$\langle\phi(0,t)\phib(p=0,t=0)\rangle$, whose Laplace transform is
$s^{-1}(1+hI_d(s^{1/2}))^{-1}$, of which the divergence for $d=2$ cannot be
removed by the renormalisation of $h$. We therefore define $\phi_R(0,t)=
Z_0\phi(0,t)$ to remove this factor, where
$Z_0=1+(B(d)/\epsilon)h\kappa^{-\epsilon}$.

Consider now $C_R^{(1)}(t,g_R,\el_R,\kappa)=\langle\phi_R(0,t)\rangle.$ This
satisfies a renormalisation group equation
$$
\left(\kappa{\partial\over\partial\kappa}+\beta_g(g_R){\partial\over
\partial g_R}+\beta_\el(\el_R){\partial\over\partial\el_R}
-\gamma_0(g_R)\right)C_R^{(1)}=0
$$
where $\gamma_0=(\kappa\partial/\partial\kappa)\ln Z_0=-B(d)g_R$.
The solution as $t\to\infty$, for $d<2$, is
$$
C_R^{(1)}(t,g_R,\el_R,\kappa)\sim
(\kappa^2t)^{-d/2}
(\kappa^2t)^{-\epsilon/2}C_R^{(1)}(\kappa^{-2},g^*,\tilde\el(t),\kappa)
\eqno(RG)
$$
where $g^*=\epsilon/B(d)$, and $\tilde\el(t)$ is the running coupling
$\el_R$. The first prefactor on the right hand side comes from the
canonical scaling dimension of $\phi$, the second from the anomalous
dimension $\gamma_0(g^*)$.

For $k>2$, there is a regime where $d_c(k)<d<2$. In this case,
$\lambda$ is irrelevant and $\tilde\lambda\sim t^{\epsilon'/d_c(k)}$,
with $\epsilon'<0$. The correlation function of the right hand side
of \(RG) is then given by the sum of tree diagrams,
equivalent to solving the inhomogeneous rate equation \(RE)
(with $\phi$ replacing $n$)
with the boundary condition
$\kappa^\epsilon g^*\phi(0)=\lim_{r\to0}S_dr^{d-1}
\partial\phi/\partial r$, which comes from varying with respect to
$\phib(0)$ and integrating by parts. $S_d$ is the area of a unit
$d$-dimensional sphere, and $\tilde\lambda=\tilde\el\kappa^\epsilon$.
The solution is proportional to
${\tilde\lambda}^{-d_c(k)/2}$, so that
$C_R^{(1)}(t)\sim1/t^{1+\epsilon'/2}$. As before,
the higher cumulants are irrelevant for $d>d_c(k)$, so we obtain the
stretched exponential result for $P(t)$ given in the summary below.
At $d=2$, the prefactor $(\kappa^2t)^{-\epsilon/2}$ is replaced by
$(\ln(\kappa^2t))^{-1}$. This results in an extra $(\ln t)^{-1}$
factor in the exponent.

When $d<d_c(k)$, $\tilde\el$ also flows towards a non-trivial fixed
point $\el^*=O(\epsilon')$, so that the correlation function on the
right hand side of \(RG) is asymptotically independent of $t$. The
prefactors combine to give a simple
$1/t$ dependence, which integrates to give $\ln t$. The amplitude of
this term is given, to leading order in $\epsilon'$ by setting
$\el=\el^*$ in the solution of \(RE), which gives a {\it universal}
amplitude of $O({\epsilon'}^{-d_c(k)/2})$.
Corrections to this come from loop
corrections to the right hand side, and, more importantly, from the
higher order cumulants, which satisfy similar renormalisation group
equations and whose integrals all scale like $\ln t$. However,
their amplitudes are suppressed for small $\epsilon'$ by powers of
${\epsilon'}^{(k-2)/(k-1)}$. For $d=d_c(k)$, $\tilde\el(t)$ flows to zero
like $(\ln t)^{-1}$, so that $C^{(1)}\sim (1/t)(\ln t)^{1/(k-1)}$, with
the higher order cumulants being suppressed only by powers of $\ln t$.

To summarise the different cases when $k>2$, we have
$$
P(t)=\cases{{\rm e}^{-\const t^{1-d_c(k)/2}},&$d>2$;\cr
{\rm e}^{-\const t^{1-d_c(k)/2}/\ln t},&$d=2$;\cr
{\rm e}^{-\const t^{(d-d_c(k))/2}},&$2>d>d_c(k)$;\cr
{\rm e}^{-\const(\ln t)^{k/(k-1)}},&$d=d_c(k)$;\cr
t^{-\theta},&$d<d_c(k)$, with $\theta=O({\epsilon'}^{-d_c(k)/2})$.\cr}
$$

Turning now to $k=2$, the case $d>2$ has already been discussed.
For $d<2$ the same arguments as for $k>2$ show that $P(t)\sim
t^{-\theta}$, with a universal exponent. However, the dependence on
$\epsilon$ is now of a different form. To leading order,  we may
solve the rate equation to calculate the right hand side of \(RG), setting
$\tilde\el=\el^*$ and $g^*=0$. (Higher order terms in $g^*$ are higher
order in $\epsilon$.) This means that the amplitude is, to
leading order, that of the bulk density\refto{BLEE}
$\langle\phi(t)\rangle\sim1/(4\pi\epsilon)t$. In addition, there is a
factor of $hZ_0^{-1}\sim 2\pi\epsilon$ in relating
$h\langle\phi(0,t)\rangle$ to $C_R^{(1)}(t)$. Thus the factors of
$\epsilon$ cancel, and we find
$$
\theta=\frac12+O(\epsilon)
\eqno(TH)
$$
The case $d=2$ is the most interesting, since both $h$ and $\lambda$ are
marginally irrelevant there. In this case the prefactor
behaves as $(4\pi/h)(\ln t)^{-1}$, while $\phi(0,t)$ (to leading order
in $g_R$) behaves as\refto{BLEE} $(1/8\pi)(\ln t/t)$. Thus we get a
competition between the two running couplings which results
in a power behaviour for $P(t)$, with $\theta=\frac12$,
consistent with its
limit as $\epsilon\to0$. Although \(TH) has the appearance of a
conventional mean field result with $O(\epsilon)$ corrections below
$d=d_c=2$, this is not the case: the value of $\theta$ for $d=2$ comes
about by a subtle cancellation of fluctuation effects, and the
exponent, for $d>d_c$, is not universal.
In addition, for $d=2$, the corrections to $C^{(1)}(t)$ are suppressed
by powers of $\ln t$ only. Thus the leading corrections to
$\ln P(t)$ are proportional to $\int^t(1/t'\ln t')dt'\sim\ln\ln t$.
This will give a logarithmic prefactor multiplying the power law
$t^{-1/2}$. Unfortunately, the calculation of the exponent of this
logarithm requires a two-loop calculation which is more difficult.

In principle, it is possible to compute higher order terms in the
$\epsilon$-expansion \(TH). However, a similar expansion\refto{BLEE}
for the bulk density
amplitude does not appear to extrapolate well to $d=1$. The $O(\epsilon)$
corrections to \(TH), which come from the second cumulant in \(CUM),
are expected
to be negative, consistent with the result of Derrida {\it et
al.}\refto{DBG}, who find that $\theta\approx0.37$ for $d=1$.
It is also straightforward to extend the analysis of the case $k=2$ to
include the reaction $A+A\to A$. If the rate for this second process is
$\lambda'$, the effect is to change the interaction part of the action
$S$ to $(2\lambda+\lambda')\phib\phi^2+(\lambda+\lambda')\phib^2\phi^2$.
This may be brought back to the standard form\refto{BLEE} by rescaling
$\phi=\xi\phi'$, $\phib=\xi^{-1}\phib'$, where
$\xi=2(\lambda+\lambda')/(2\lambda+\lambda')$. The result is that
the density amplitude for $d<2$, and therefore the exponent $\theta$
to lowest order, is modified by this factor of $\xi$.\refto{BLEE}
In one dimension, the reaction $A+A\to A$ occurs in the domain wall
dynamics of the $q$-state Potts model with $q\not=2$, with
$\lambda'/\lambda=q-2$.  Thus, to leading
order in $\epsilon$, the power is modified to
$$
\theta={q-1\over q}+O(\epsilon)
$$
so that, at least to this order, $\theta$ increases with $q$ as found for
$d=1$.\refto{DBG} The $q$-dependence is, however, more complicated
for the higher order terms.

The author thanks A.~J.~Bray and B.~Derrida for communicating their
work prior to publication, and B.~P.~Lee for conversations on the
renormalisation group approach and a reading of the manuscript.
This work was supported by a grant from the EPSRC.

\references

\refis{BLEE} B.~P.~Lee, \journal J. Phys. A, 27, 2633, 1994.

\refis{DBG} B.~Derrida, A.~J.~Bray and C.~Godr\`eche, \journal J. Phys.
A, 27, L357, 1994.

\refis{PEL} L.~Peliti, \journal J. Phys. A, 19, L365, 1986; \journal
J. Physique, 46, 1469, 1985.

\refis{DOI} M.~Doi, \journal J. Phys. A, 9, 1465, 1976; \journal
J. Phys. A, 9, 1479, 1976.

\refis{A} K.~Kang and S.~Redner, \journal Phys. Rev. A, 32, 435, 1985;
V.~Kuzovkov and E.~Kotomin, \journal Rep. Prog. Phys., 51,
1479, 1988

\endreferences

\endit